\documentclass[reprint,superscriptaddress,showpacs,amsmath, amssymb,aps,pra]{revtex4-1}
\usepackage{graphicx}
\usepackage{dcolumn}
\usepackage{bm}
\usepackage{subfigure}
\usepackage{hyperref}
\hypersetup{colorlinks=true, citecolor=blue, urlcolor=blue, linkcolor=blue}
\usepackage[ruled]{algorithm2e}
\begin{document}
\title{Quantum Anomaly Detection with Density Estimation and Multivariate Gaussian Distribution}
\author{Jin-Min Liang}
\affiliation{College of Science, China University of Petroleum, 266580 Qingdao, P.R. China.}
\author{Shu-Qian Shen}
\email{sqshen@upc.edu.cn.}
\affiliation{College of Science, China University of Petroleum, 266580 Qingdao, P.R. China.}
\author{Ming Li}
\affiliation{College of Science, China University of Petroleum, 266580 Qingdao, P.R. China.}
\affiliation{Max-Planck-Institute for Mathematics in Sciences, 04103 Leipzig, Germany}
\author{Lei Li}
\affiliation{College of Science, China University of Petroleum, 266580 Qingdao, P.R. China.}
\date{\today}
\begin{abstract}
We study quantum anomaly detection with density estimation and multivariate Gaussian distribution. Both algorithms are constructed using the standard gate-based model of quantum computing. Compared with the corresponding classical algorithms, the resource complexities of our quantum algorithm are logarithmic in the dimensionality of quantum states and the number of training quantum states. We also present a quantum procedure for efficiently estimating the determinant of any Hermitian operators $\mathcal{A}\in\mathcal{R}^{N\times N}$ with time complexity $O(poly\log N)$ which forms an important subroutine in our quantum anomaly detection with multivariate Gaussian distribution. Finally, our results also include the modified quantum kernel principal component analysis (PCA) and the quantum one-class support vector machine (SVM) for detecting classical data.
\end{abstract}
\maketitle
\section{Introduction}
Anomaly detection (AD) is a pivotal topic in statistics that has been researched within diverse areas and application domains, such as fraud detection for credit cards, insurance or health care, intrusion detection for cyber-security, fault detection in safety critical systems, and military surveillance for enemy activities \cite{Laurikkala2000,Aleskerov1997}. The importance of AD is due to the fact that anomalous data indicate significant information. For example, an anomalous MRI image may mean presence of malignant tumors \cite{Spence2003}. Traditional anomaly detection techniques focus on detecting anomalous data that differ from normal (or clean) data patterns. Over time, a variety of anomaly detection techniques have been developed in several research communities. One of the most common machine learning algorithms for AD is based on density estimation method by estimating the probability density function \cite{Breunig2000,Ramaswamy2000,Lazarevic2005,Shyu2003}. In 2009, the authors presented a first attempt to evaluate two previously proposed methods for statistical anomaly detection in sea traffic, namely the Gaussian mixture model and the adaptive kernel density estimator \cite{Laxhammar2009}. Erfani \emph{et al.} proposed a hybrid model for unsupervised anomaly detection that combines a one-class support vector machine (SVM) and a deep belief network (DBN) \cite{Erfani2016}. Carrera \emph{et al.} presented a novel approach for detecting anomalous structures in images by learning a convolutional sparse model that describes the local structures of normal images \cite{Carrera2015}. Especially for point anomalies, Hoffmann studied kernel PCA for novelty detection in 2007 \cite{Hoffmann2007}. Due to the fact that abnormalities are very rare, we could not use the traditional classification method to learn an anomaly model described by the fault conditions. AD offered a solution to this situation by modeling a normal data set and using a distance measure and a threshold for determining abnormality \cite{Markou2003}.

In parallel, we are witnessing the influence of quantum information processing on its classical counterpart. Following the discovery of quantum algorithms for factoring \cite{Shor1994}, database searching \cite{Grover1996} and quantum matrix inverse \cite{HHL2009}, decades of work have shown that quantum algorithms have the capability of outperforming existing classical methods. These algorithms have recently been employed in machine learning and form an interdisciplinary field called quantum machine learning (QML). In particular, it has proved useful in regression and classification problems that appear in machine learning \cite{QSVM2014,Liu2018,Schuld2016,Cong2016,Lloyd2018,BDuan2019}.

A natural question arises of how to design a quantum machine learning algorithm for detecting anomalies in a quantum computer. Although some previous results have been reported \cite{QCP2017,QCP2018,Liu2018}, AD still requires more investigation from a quantum mechanics perspective. Quite recently, Sent\'{i}s \emph{et al.} introduced the quantum change point and devised online strategies for identifying a change point in a stream of quantum particles allegedly prepared in identical states \cite{QCP2017,QCP2018}, which is a quantum version of identifying changes in a sequence of random variables. Liu \emph{et al.} designed two quantum algorithms for quantum anomaly detection, quantum kernel PCA and quantum one-class SVM, proved that these two quantum algorithms can be performed using resources logarithmic in the dimensionality of quantum states \cite{Liu2018}.

In the present study, we design two quantum anomaly detection algorithms based on density estimation and multivariate Gaussian distribution. Given a training set of $M$ quantum states which have $d$ dimensional features, we can identify which one varies significantly from the average. We show how a training set can be modeled by using a Gaussian distribution and how the model can be used for anomaly detection. Our algorithm can be achieved using resources only logarithmic in the dimensionality of quantum states and the number of training quantum states. The organization of the paper is as follows. In Sec. II, we review the classical anomaly detection algorithm with density estimation and multivariate Gaussian distribution. In Secs. III and IV, we introduce the key quantum anomaly detection algorithm with density estimation and multivariate Gaussian distribution in detail. We discuss the generalization of quantum kernel PCA and quantum one-class SVM to tackle classical discrete data anomaly in Sec. V. A summary and discussions are included in Sec. VI.
\section{Anomaly detection algorithm}
In this section, we provide the necessary background information to understand this paper. An anomaly detection algorithm aims to construct a data model which normal data obeys whereas anomalous data do not. Anomalies are patterns in data that do not meet a well defined notion of normal behavior \cite{Chandola2009}. For instance, Fig. 1 illustrates anomalies in a simple two-dimensional data set. The data set has two normal regions, $R_1$ and $R_2$, since most observations lie in these two regions. Points that are sufficiently far away from the regions, e.g., points $p_1$ and $p_2$, and points in region $R_3$, are anomalies. Typically, one can estimate a number of parameters characterized by the data model of a random variable. Parametric approaches make an assumption that data distributions are Gaussian in nature and they can be modeled statistically based on data means and covariance \cite{Markou2003}. We will use an anomaly detection algorithm with density estimation and multivariate Gaussian distribution, described below, as a foundation to present a quantum anomaly detection algorithm.
\begin{figure}[htbp]
\centering
\includegraphics[width=3in]{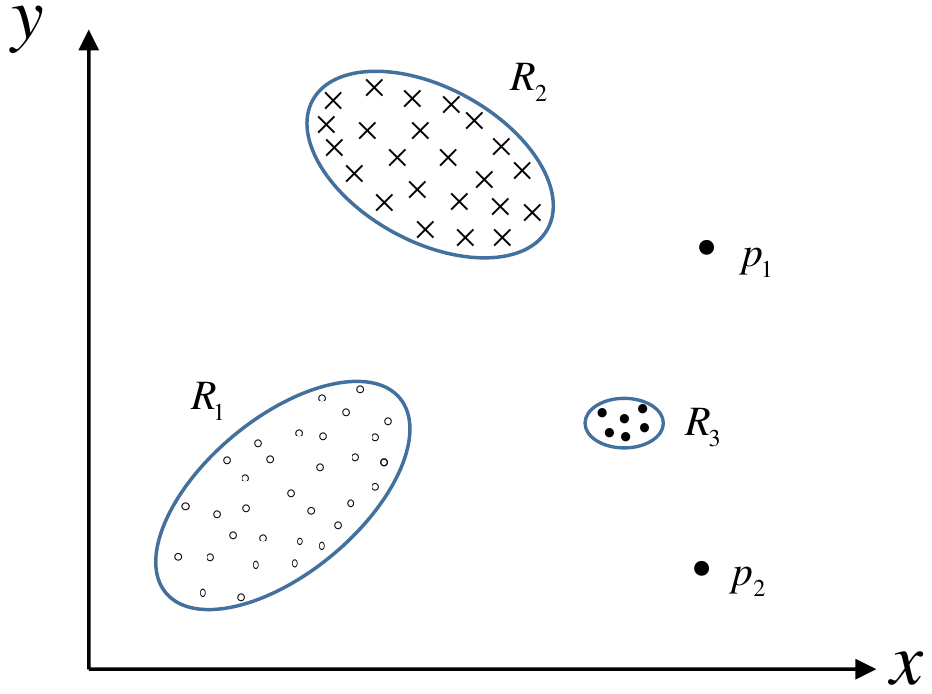}
\caption{A simple example of anomalies in a two-dimensional data set.}
\end{figure}
\subsection{Anomaly detection algorithm with density estimation}
Suppose we are given a training set $\{x^{i}\}_{i=1}^{M}$ with $d$ features and $M$ samples which are expected to follow a Gaussian distribution. The algorithm aims to create a statistical model $p(x)$ from the training set by assuming each feature follows a Gaussian distribution and then identifying whether the new datum $x$ is abnormal or not. Then the overall procedure is depicted as follows:

1) Fit density parameters $\mu_1,\cdots,\mu_d,\sigma_1^2,\cdots,\sigma_d^2$, where
$$\mu_j=\frac{1}{M}\sum_{i=1}^{M}x_j^i, \sigma_j^2=\frac{1}{M}\sum_{i=1}^{M}(x_j^i-\mu_j)^2.$$

2) Given a new example $x$, compute $p(x)$,
\begin{equation}
p(x)=\prod_{j=1}^{d}p(x_j;\mu_j,\sigma_j^2)=\prod_{j=1}^{d}\frac{1}{\sqrt{2\pi}\sigma_j}e^{-\frac{(x_j-\mu_j)^2}{2\sigma_j^2}}.
\end{equation}

3) If $p(x)<\epsilon$, one can flag this datum as an anomaly, otherwise, it is a normal datum.

The parameter $\epsilon$ is some threshold probability value depending on how sure we want to be. The most common method for selecting a suitable threshold is K-fold cross validation the same as other prediction tasks \cite{Kohavi1995,Rodriguez2010}. For simplicity, we take the following computation based on Eq. (1):
\begin{equation}
\begin{aligned}
&\ln p(x)\\
&=\ln\prod_{j=1}^{d}\frac{1}{\sqrt{2\pi}\sigma_j}e^{-\frac{(x_j-\mu_j)^2}{2\sigma_j^2}}
=\sum_{j=1}^{d}\ln\frac{1}{\sqrt{2\pi}\sigma_j}e^{-\frac{(x_j-\mu_j)^2}{2\sigma_j^2}}\\
&=-\frac{d}{2}\ln2\pi-\sum_{j=1}^{d}\ln\sigma_j-\sum_{j=1}^{d}\frac{(x_j-\mu_j)^2}{2\sigma_j^2}.
\end{aligned}
\end{equation}
\subsection{Anomaly detection algorithm with multivariate Gaussian distribution}
The multivariate Gaussian distribution is different from the Gaussian distribution. This algorithm attempts to develop a model $p(x)$ all in one process, instead of modeling each feature separately as in Eq. (1). For a training set $\{x^{i}\}_{i=1}^{M}$ with $d$ features and $M$ samples, we fit model $p(x)$ by setting
\begin{equation}
\mu=\frac{1}{M}\sum_{i=1}^{M}x^{i},\quad C=\frac{1}{M}\sum_{i=1}^{M}(x^{i}-\mu)(x^{i}-\mu)^{T}.
\end{equation}
Given a new example $x$, compute
\begin{equation}
p(x)=\frac{1}{(2\pi)^{\frac{d}{2}}|C|^{\frac{1}{2}}}
e^{-\frac{1}{2}(x-\mu)^{T}C^{-1}(x-\mu)},
\end{equation}
where $|C|$ represents the determinant of matrix $C$. If $p(x)<\epsilon$ we can flag this datum as an anomaly, otherwise, it is a normal datum.
\section{Quantum Anomaly detection algorithm with Density Estimation}
In this section, we will develop a quantum anomaly algorithm with density estimation. For classical data, our schemes require the method of encoding classical information into the amplitude of a quantum system and then applying our quantum anomaly detection algorithms directly. In other words, we encode a $2^n-$ dimensional vector $\vec u=\{u_0,u_1,\cdots,u_{2^n-1}\}$ into the $2^n$ amplitudes $u_0,u_1,\cdots,u_{2^n-1}$ of an $n$-qubit quantum system, $|u\rangle=\sum_{i=0}^{2^n-1}u_i|i\rangle$, where $\{|i\rangle\}$ is the computational basis \cite{Schuld2016}. Thus, in what follows, we only consider quantum pure state anomaly detection. In that sense, our algorithms can detect an anomaly in both classical data and quantum states generated from quantum devices.

Suppose we are given access to the unitaries $\{U_{i}\}_{i=0}^M$ to obtain quantum states $\{|x^i\rangle\}_{i=0}^{M}$, where $U_{i}|0\cdots0\rangle=|x^i\rangle$. We are given a set of $M$ \textit{normal} quantum training states $\{|x^i\rangle\}_{i=1}^{M}$ and the test quantum state $\{|x^0\rangle\}$. Our task is to detect how anomalous the state $|x^0\rangle$ is compared to the normal states. We assume the runtime of our algorithm is dominated by the number of quantum gates. First, we define the analogous mean state of the training quantum state by
\begin{equation}
|\mu\rangle=\frac{1}{N_{\mu}}\sum_{i=1}^{M}|x^i\rangle
=\frac{1}{N_{\mu}}\sum_{i=1}^{M}\sum_{j=1}^{d}x_j^i|j\rangle,
\end{equation}
where $x_j^i$ denotes the $j$th element of state $|x^i\rangle$.  The normalization coefficient $N_{\mu}^2=\sum_{k,l=1}^{M}\langle x^k|x^l\rangle$ can be found using $O(\log M)$ resources. Here, we show a method for preparing the mean state. We can apply the unitary operation $I\otimes H^{\log M}$ on the given initial state $|\Psi_1\rangle=\frac{1}{\sqrt{M}}\sum_{i=1}^M|x^i\rangle|i\rangle$ to obtain
\begin{equation}
\begin{aligned}
|\psi_\mu\rangle=\frac{1}{M}\sum_{i=1}^M|x^i\rangle\sum_{j=1}^M(-1)^{i\cdot j}|j\rangle,
\end{aligned}
\end{equation}
where $i\cdot j$ is the bitwise inner product of $i$ and $j$, modulo 2. Specifically, we perform a Hadamard gate $H$ on each of the qubits of the second register. Then by making a projective measurement $|0\cdots0\rangle\langle0\cdots0|$ on the second register we can recover the mean state $|\mu\rangle|0\cdots0\rangle$. The success probability $P_\mu$ of this measurement is
\begin{equation}
\begin{aligned}
P_\mu&=tr(|0\cdots0\rangle\langle0\cdots0|H^{\otimes \log M}|\Psi_1\rangle\langle\Psi_1|H^{\otimes \log M})\\
&=\frac{1}{M^2}\sum_{k,l=1}^M\langle x^k|x^l\rangle.
\end{aligned}
\end{equation}
Obviously, the use of an $O(\log M)$ Hadamard gates means that our gate resource count is $O(\log M)$.
\begin{algorithm}[]
\caption{Efficiently computing $\sum_{j=1}^d\frac{(x_j^0-\mu_j)^2}{2\sigma_j^2}$}
\LinesNumbered
\emph{Step 1.} Prepare the quantum state $\sum_{j=1}^d|\chi_j\rangle|\chi_j^0\rangle|j\rangle.$\\
\emph{Step 2.} Add an ancilla qubit $|0\rangle$ and perform a controlled unitary operator $R_1$ to obtain
$$\sum_{j=1}^d|\chi_j\rangle|\chi_j^0\rangle|j\rangle
\Bigg(\frac{\chi_j^0}{\chi_j}|0\rangle+\sqrt{1-\Bigg(\frac{\chi_j^0}{\chi_j}\Bigg)^2}|1\rangle\Bigg).$$\\
\emph{Step 3.} Uncompute the second and third registers. The system state is
$$\sum_{j=1}^d|j\rangle
\Bigg(\frac{\chi_j^0}{\chi_j}|0\rangle+\sqrt{1-\Bigg(\frac{\chi_j^0}{\chi_j}\Bigg)^2}|1\rangle\Bigg).$$\\
\emph{Step 4.} Measure the observable $M_1=I\otimes|0\rangle\langle0|$ and the expectation
$\langle M_1\rangle=\sum_{j=1}^d\Bigg(\frac{\chi_j^0}{\chi_j}\Bigg)^2.$
\end{algorithm}

Next, we show a method for preparing the state $|\chi\rangle=\frac{1}{N_{\chi}}\sum_{j=1}^d|\chi_j\rangle|j\rangle=\frac{1}{N_{\chi}}\sum_{j=1}^d\sum_{i=1}^M(x_j^i-\mu_j)|i\rangle|j\rangle$ corresponding to the $x_j-\mu_j$ in classical anomaly detection with density estimation, where $N_{\chi}^2=\sum_{k,l=1}^d\langle\chi_k|\chi_l\rangle$. We rewrite the mean state $|\mu\rangle$ as $|\mu\rangle=\frac{1}{N_{\mu}}\sum_{j=1}^d\mu_j|j\rangle$, where $\mu_j=\sum_{i=1}^Mx_j^i$. Suppose we are given the control unitaries to create the following superpositions of training states
\begin{equation}
\begin{aligned}
|\Psi_2\rangle=\frac{|0\rangle\sum_{j=1}^d|x^i\rangle|j\rangle+|1\rangle|\mu\rangle\sum_{i=1}^M|i\rangle}{\sqrt{2}}.
\end{aligned}
\end{equation}
Then we apply a Hadamard gate on qubit of the first register to obtain
\begin{equation}
\begin{aligned}
\frac{1}{2}
\Bigg (|0\rangle\sum_{j=1}^d\sum_{i=1}^M(x_j^i+\mu_j)|i\rangle|j\rangle+|1\rangle\sum_{j=1}^d\sum_{i=1}^M(x_j^i-\mu_j)|i\rangle|j\rangle\Bigg ).
\end{aligned}
\end{equation}
As the final step measure the first register in $|1\rangle$, the remaining qubits collapse into the state
\begin{equation}
\begin{aligned}
\sum_{j=1}^d\sum_{i=1}^M(x_j^i-\mu_j)|i\rangle|j\rangle=\sum_{j=1}^d|\chi_j\rangle|j\rangle.
\end{aligned}
\end{equation}
The success probability $P_\chi$ of this measurement is
\begin{equation}
\begin{aligned}
P_\chi&=tr(|1\rangle\langle1|H|\Psi_2\rangle\langle\Psi_2|H)\\
&=\sum_{k,l=1}^d\langle\chi_k|\chi_l\rangle.
\end{aligned}
\end{equation}
In this process, our gate resource count is $O(1)$. The same trick as above, for the test quantum state $|x^0\rangle$ and the mean state $|\mu\rangle$, we can easily prepare state $\sum_{j=1}^d\sum_{i=1}^M(x_j^0-\mu_j)|i\rangle|j\rangle=\sum_{j=1}^d|\chi_j^0\rangle|j\rangle$, where $|\chi_j^0\rangle=\sum_{i=1}^M(x_j^0-\mu_j)|i\rangle$.

\begin{algorithm}[]
\caption{Efficiently computing $\sum_{j=1}^d\ln \delta_j$}
\LinesNumbered
\emph{Step 1.} According to Eq.(10), we obtain the quantum state $\sum_{j=1}^d|\chi_j\rangle|j\rangle.$\\
\emph{Step 2.} Add an ancilla qubit $|0\rangle$ and perform a controlled unitary operator $R_2$ to obtain
$$\sum_{j=1}^d|\chi_j\rangle|j\rangle
\Bigg(\ln \chi_j|0\rangle+\sqrt{1-2\ln \chi_j}|1\rangle\Bigg).$$\\
\emph{Step 3.} Uncompute the second registers . The system state is
$$\sum_{j=1}^d|j\rangle
\Bigg(\ln \chi_j|0\rangle+\sqrt{1-2\ln \chi_j}|1\rangle\Bigg).$$\\
\emph{Step 4.} Measure the observable $M_2=I\otimes|0\rangle\langle0|$ and the expectation
$\langle M_2\rangle=2\sum_{j=1}^d\ln \chi_j.$
\end{algorithm}

With these preparations, we present Algorithm 1 for calculating the $\sum_{j=1}^d\frac{(x_j^0-\mu_j)^2}{2\sigma_j^2}$. In step 2, the unitary operator $R_1$ transform $|0\rangle$ to $\frac{\chi_j^0}{\chi_j}|0\rangle+\sqrt{1-\Bigg(\frac{\chi_j^0}{\chi_j}\Bigg)^2}|1\rangle$ controlled by the states $|\chi_j\rangle$ and $|\chi_j^0\rangle$. Thus, we can easily deduce $\sum_{j=1}^d\frac{(x_j^0-\mu_j)^2}{2\sigma_j^2}=\frac{1}{2}\langle M_1\rangle$.

At the same time, we present Algorithm 2 for computing $\sum_{j=1}^d\ln \delta_j$. In step 2, the unitary operator $R_2$ transform $|0\rangle$ to
$\ln\chi_j|0\rangle+\sqrt{1-2\ln \chi_j}|1\rangle$ controlled by the states $|\chi_j\rangle$. Thus, we can easily deduce $\sum_{j=1}^d\ln \delta_j=\frac{1}{2}\langle M_2\rangle$. Then, given a probability threshold $\epsilon$, these two algorithms allow us to detect anomalies based upon Algorithm 3. If
$$\ln p(|x^0\rangle)=-\frac{d}{2}\ln2\pi-\frac{1}{2}\Bigg(\langle M_1\rangle+\langle M_2\rangle\Bigg)<\ln\epsilon,$$
one can flag $|x^0\rangle$ as a normal quantum state, otherwise, $|x^0\rangle$ is an abnormal quantum state. The overall resources complexity of Algorithm 3 is $O(\log M)$.
\begin{algorithm}[]
\caption{Quantum anomaly detection with density estimation}
\LinesNumbered
\emph{Input.}  The quantum state training set $\{|x^i\rangle\}_{i=1}^{M}$, the test quantum state $|x^0\rangle$ and a suitable threshold $\epsilon$.\\
\emph{Step 1.} Apply Algorithm 1 to obtain $\langle M_1\rangle$ using quantum gate resources $O(\log M)$.\\
\emph{Step 2.} Apply Algorithm 2 to obtain $\langle M_2\rangle$ using quantum gate resources $O(1)$.\\
\emph{Step 3.} If $$\ln p(|x^0\rangle)=-\frac{d}{2}\ln2\pi-\frac{1}{2}\Bigg(\langle M_1\rangle+\langle M_2\rangle\Bigg)<\ln\epsilon,$$
then $|x^0\rangle$ is a normal quantum state. Otherwise, $|x^0\rangle$ is an abnormal quantum state.
\end{algorithm}

\section{Quantum Anomaly detection algorithm with Multivariate Gaussian Distribution}
Here, we show how to identify an anomaly with a multivariate Gaussian distribution on a quantum computer. For simplicity, we transform Eq. (4) to a fairly easy form,
\begin{equation}
p_{test}=(x-\mu)^{T}C^{-1}(x-\mu)>2\ln \Bigg((2\pi)^{\frac{n}{2}}|C|^{\frac{1}{2}}\varepsilon\Bigg).
\end{equation}
We also give two algorithms to efficiently compute the subitems $p_{test}$ and the determinant of $C$. In subsection A, we show the computation of $p_{test}$. Then, we describe the computation process of estimating the determinant of a given Hermitian positive definite operator in subsection B. Since our algorithms make use of a quantum linear systems algorithm \cite{HHL2009}, we achieve an exponential speed-up over its classical counterpart.

\subsection{Efficient computation of $p_{test}$}
 For a quantum anomaly detection algorithm with multivariate Gaussian distribution, first we define the centered state of the training quantum state by
\begin{equation}
\begin{aligned}
|z^i\rangle=\sum_{j=1}^{d}z_j^i|j\rangle,
\end{aligned}
\end{equation}
where $z_j^i$ is the $j$th component of the vector $z^i$ and $z^i=x^i-\mu$. Following the method of \cite{Cong2016}, we will assume the following oracle:
\begin{equation}
\begin{aligned}
O_1(|i\rangle|0^{\otimes \log d}\rangle)\rightarrow|i\rangle|z^i\rangle,
\end{aligned}
\end{equation}
to get the state $|z^i\rangle$ in time $O(\log d)$. Obviously, given a new example $x^0$, we can also obtain the corresponding centered state $|z^0\rangle$. The covariance matrix can be rewritten in terms of quantum states as $C=\frac{1}{M-1}\sum_{i=1}^M|z_j^i\rangle\langle z_j^i|$. It is also proportional to a density matrix $\mathcal{C}=\frac{C}{tr(C)}$, where $tr(C)=\frac{1}{M-1}\sum_{i=1}^M\sum_{j=1}^dz_j^i(z_j^i)^{*}$. Since $\mathcal{C}$ is Hermitian operator, we can use the method of \cite{QPCA2014} to exponentiate this operator. Assuming the operator $\mathcal{C}$ has spectral decomposition $\mathcal{C}=\sum_{k=1}^d\lambda_{k}|u_k\rangle|\langle u_k|$, we can obtain
\begin{equation}
\begin{aligned}
p_{test}&=\langle z^0|C^{-1}|z^0\rangle=\frac{\langle z^0|\mathcal{C}^{-1}|z^0\rangle}{tr(C)}\\
&=\frac{1}{tr(C)}\sum_{j=1}^d\frac{\beta_j^2}{\lambda_{j}}.
\end{aligned}
\end{equation}
At the same time, we decompose $|z^0\rangle$ in the eigenbasis of $\mathcal{C}$.
To avoid exponential small eigenvalues, we define an effective condition number $\kappa$ and take into account only eigenvalues in the interval $[1/\kappa,1]$. Here, we give a quantum Algorithm 4 for tackling this problem based on the phase estimation \cite{PE1996}. The expectation value for the measurement in the final state $|\Psi_3\rangle$ is $\langle M_3\rangle=\sum_j\frac{\beta_j^2}{\lambda_j}.$ Thus, we can easily deduce the value of $p_{test}=\langle M_3\rangle$.

\begin{algorithm}[]
\caption{Efficiently computing $p_{test}$}
\LinesNumbered
\emph{Step 1.} Prepare the quantum states $|0\rangle$ and $|z^0\rangle$ in registers $A,B$\\
\emph{Step 2.} Since the operator $\mathcal{C}$ is Hermitian positive semidefinite, use the well-known technique of phase estimation \cite{PE1996} to obtain the eigenvalue/eigenvector, in time $O(\log d)$:
$$
\sum_{j=1}^d\beta_j|\lambda_j\rangle_A|u_j\rangle_B, \beta_j=\langle z^0|u_j\rangle.
$$\\
\emph{Step 3.} Add a joint qubit $|0\rangle$ and invert the square root of eigenvalues to obtain
$$
\sum_{j=1}^d\beta_j|\lambda_j\rangle_A|u_j\rangle_B
(\frac{1}{\sqrt{\lambda_j}}|0\rangle+\sqrt{1-\frac{1}{\lambda_j}}|1\rangle).
$$\\
\emph{Step 4.} Uncompute the output of the phase estimation and obtain the state
$$
|\Psi_3\rangle=\sum_{j=1}^d\beta_j|u_j\rangle_B
(\frac{1}{\sqrt{\lambda_j}}|0\rangle+\sqrt{1-\frac{1}{\lambda_j}}|1\rangle)
$$\\
\emph{Step 5.} Measure the observable $M_3=I_B\otimes|0\rangle\langle0|$ and the expectation $\langle M_3\rangle=\sum_{j=1}^d\frac{\beta^2}{\lambda_j}.$
\end{algorithm}

\subsection{Efficient estimation $\ln|C|^{\frac{1}{2}}$}
In this subsection, we will describe a method to estimate the determinant of a Hermitian positive definite operator, which plays key roles in machine learning and optimization \cite{Bishop2006}. However, the determinant of matrices has been less investigated from a quantum information perspective. For a Hermitian operator $\mathcal{B}\in\mathcal{R}^{n\times n}$ with spectral decomposition $\mathcal{B}=\sum_{i=1}^n\lambda_iu_iu_i^{\dag}$, the determinant of $\mathcal{B}$ is given by
\begin{equation}
\begin{aligned}
|\mathcal{B}|=\Pi_{i=1}^n\lambda_i=e^{\sum_{i=1}^d\ln\lambda_i}.
\end{aligned}
\end{equation}
Thus, we only need to achieve the exponential of $\sum_{i=1}^d\ln\lambda_i.$ Motivated by the Harrow-Hassidim-Lloyd algorithm \cite{HHL2009}, we design Algorithm 5 to estimate the determinant of Hermitian positive definite operator. Next, we would like to establish error bounds on the final result in Algorithm 5. We first define the map
\begin{equation}
\begin{aligned}
|g(\lambda)\rangle:=\ln\lambda|0\rangle+\sqrt{1-2\ln\lambda}|1\rangle.
\end{aligned}
\end{equation}
This allows us to obtain the logarithm of the eigenvalue. We will make use of the following lemma to bound the error.

\textbf{Lemma 1.} \emph{The map $\lambda\mapsto|g(\lambda)\rangle$ is $O(\kappa^2)$-Lipschitz, meaning that for any $\lambda_i\neq\lambda_j$,
\begin{equation}
\begin{aligned}
\||g(\lambda_i)\rangle-|g(\lambda_j)\rangle\|\leq c\kappa^2|\lambda_i-\lambda_j|.
\end{aligned}
\end{equation}
for some $c=\mathcal{O}(1)$}.

\emph{Proof.} Since $\lambda\mapsto|g(\lambda)\rangle$ is continuous everywhere and differentiable, it suffices to bound the norm of the derivative of $|g(\lambda)\rangle$. We consider
\begin{equation}
\begin{aligned}
\frac{d}{d\lambda}|g(\lambda)\rangle=\frac{1}{\lambda}|0\rangle-\frac{1}{\lambda\sqrt{1-2\ln\lambda}}|1\rangle.
\end{aligned}
\end{equation}
Next, the norm of $\frac{d}{d\lambda}|g(\lambda)\rangle$ is
\begin{equation}
\begin{aligned}
\frac{1}{\lambda^2}+\frac{1}{\lambda^2(1-2\ln\lambda)}\leq\frac{2}{\lambda^2}=2\kappa^2.
\end{aligned}
\end{equation}
This completes the proof, with $c=2$.$\hfill\Box$

If we require result error to be of $O(\epsilon)$, we need to take the phase estimation accuracy to be $\delta=O(\frac{\epsilon}{\kappa^2})$.

Using classical computation theory, the determinant can be expressed as a sum of products of entries of the operator where each product has $N$ items and this expression grows rapidly with the size of the operator, $O(N!)$. However, the runtime of the proposed algorithm is $O(poly(\log N))$, which is an exponential improvement over its classical counterpart.

Based on these two algorithms and Eq. (12), we can flag example quantum state $|x^0\rangle$ as an anomaly if $\langle M_3\rangle>2\ln ((2\pi)^{\frac{d}{2}}|C|^{\frac{1}{2}}\varepsilon)$, otherwise it is a normal state.
\begin{algorithm}[]
\caption{Estimating the determinant of $\mathcal{B}$}
\LinesNumbered
\emph{Step 1.} Since the operator $\mathcal{B}$ is Hermitian positive semidefinite, use the the well-known technique of phase estimation \cite{PE1996} to obtain the eigenvalue/eigenvector.\\
\emph{Step 2.} Apply the map taking $|\lambda_i\rangle$ to $\ln\lambda_i|\lambda_i\rangle$ by an ancilla qubit. Once it succeeds, we go to the next step.\\
\emph{Step 3.} Uncompute the eigenvalue register and obtain the state $\sum_{i=0}^N|u_i\rangle(\ln\lambda_i|0\rangle+\sqrt{1-(\ln\lambda_i)^2}|1\rangle).$\\
\emph{Step 4.} Measure the observable $M=I_B\otimes|0\rangle\langle0|$ and the expectation $\langle M\rangle=\sum_{i=0}^N\ln\lambda_i$. Then the determinant $|\mathcal{B}|=e^{\langle M\rangle}$.
\end{algorithm}
\section{Quantum kernel PCA and one-class SVM for classical data}
In \cite{Liu2018}, the authors presented quantum kernel PCA and one-class SVM for quantum state anomaly detection. In this section, we apply the quantum kernel PCA and one-class SVM for detecting a classical data anomaly. After encoding the discrete data set $\{x^i\}_{i=1}^M$ to quantum state $\{|x^i\rangle\}_{i=1}^M$, we can directly use the quantum one-class SVM to identify the anomaly. Although quantum kernel PCA can also directly detect an anomaly for classical data, the resources required are much more than for our modified quantum kernel PCA discussed below.
\subsection{The modified quantum kernel PCA}
We can rewrite the centered quantum state as
\begin{equation}
\begin{aligned}
|z^i\rangle=\sum_{j=1}^dz_j^i|j\rangle,
\end{aligned}
\end{equation}
the same as Eq. (13) in Sec. 4 A. We can prepare the test centered state by $|z^0\rangle=\sum_{j=1}^dz_j^0|j\rangle$. The centered sampled covariance matrix is $C=\frac{1}{M-1}\sum_{i=1}^{M}|z^i\rangle\langle z^i|$. The proximity measure $f(|z^0\rangle)$ obeys $0\leq f(|z^0\rangle)\leq1$, where
\begin{equation}
\begin{aligned}
f(|z^0\rangle)=\langle z^0|I-C|z^0\rangle.
\end{aligned}
\end{equation}
In \cite{Liu2018}, the inner product can be calculated by $M$ times the modified swap test and each of that requires $O(poly(\log d))$. Thus the proximity measure $f(|z^0\rangle)$ can be estimated using resources scaling as $O(poly(M\log d))$ \cite{Liu2018}. Here, we present a modified algorithm to efficiently compute the proximity based on Algorithm 4 using resources scaling as $O(poly(\log d))$. More specifically, we use the method of \cite{QPCA2014} to obtain the eigenvalue of $C$ and rotate one qubit $|0\rangle$ to $(1-\lambda)|0\rangle+\sqrt{1-(1-\lambda)^2}|1\rangle$ controlled by $\lambda$. The time complexity of this process is $O(poly(\log d))$ independent of the number of training quantum states.

\section{Conclusion and Discussion}
In the present study, we presented quantum anomaly detection with density estimation and multivariate Gaussian distribution. For an exponentially large data set, the resources are logarithmic in the dimensionality of quantum states and the number of training quantum states. Moreover, We also developed a quantum subroutine for efficiently estimating the determinant of any Hermitian operator $\mathcal{A}\in\mathcal{R}^{N\times N}$ with time complexity $O(\log(N))$. This algorithm might be a subroutine in machine learning and optimization. Finally, our work generalized the result given in \cite{Liu2018} to tackle a classical data anomaly while the authors had only considered quantum anomaly detection for quantum states generated from quantum devices.

Although we have seen the feasibility of detecting an anomaly on a quantum computer, some open questions still require further study in the future. For example, it would be interesting to study how to select a suitable threshold $\epsilon$ efficiently on a quantum computer. Finally, both algorithms are described in the standard gate-based model of quantum computing.  In 2007, Aharonov \emph{et.al} stated that adiabatic quantum computation \cite{Farhi2000} is as powerful as conventional quantum computation \cite{Aharonov2007}. Thus, it would be interesting to consider other alternative quantum anomaly detection algorithms with adiabatic quantum computing.

\textit{Acknowledgments} We are very grateful to the referee and the editor for their constructive suggestions. The authors thank Dr. Bo-Jia Duan for her invaluable suggestions. This work is supported by the Natural Science Foundation of Shandong Province (ZR2016AM23) and the Fundamental Research Funds for the Central Universities (18CX02035A, 18CX02023A).

\end{document}